\documentclass[aps,pra,showpacs,showkeys,superscriptaddress,twocolumn]{revtex4-1}
\usepackage[colorlinks,linkcolor=blue,anchorcolor=blue,citecolor=blue,urlcolor=blue,breaklinks=true]{hyperref}
\usepackage{graphicx}
\usepackage{amsmath}
\usepackage{amssymb}
\usepackage{slashed}
\usepackage{latexsym}
\usepackage{epsfig}
\usepackage{amsbsy}
\usepackage{array}
\usepackage{amssymb}
\usepackage{setspace}
\usepackage{bm}
\usepackage{lipsum}
\usepackage{mathrsfs}
\usepackage{float}
\usepackage{color}
\usepackage{graphicx}
\usepackage{subfigure}
\usepackage[T1]{fontenc}
\usepackage{mathptmx}
\DeclareMathAlphabet{\mathcal}{OMS}{cmsy}{m}{n}
\DeclareSymbolFont{largesymbols}{OMX}{cmex}{m}{n}

\begin{document}
\author{Zheng Zhang}
\email{zzhang@chenwang.nju.edu.cn}
\affiliation{Department of physics, Nanjing University, Nanjing 210093, China}
\author{Tong Zhao}
\email{zhao708@purdue.edu}
\affiliation{Department of physics, Nanjing University, Nanjing 210093, China}
\author{Y.-H Xia}
\affiliation{Department of physics, Nanjing University, Nanjing 210093, China}
\affiliation{Nanjing Proton Source Research and Design Center, Nanjing 210093, China}
\author{Hongshi Zong}
\email{zonghs@nju.edu.cn}
\affiliation{Department of physics, Nanjing University, Nanjing 210093, China}
\affiliation{Nanjing Proton Source Research and Design Center, Nanjing 210093, China}
\affiliation{Joint Center for Particle, Nuclear Physics and Cosmology, Nanjing 210093, China}
\date{\today}

\title{Geometry Effect on Black Body Radiation with Different Boundary Conditions}

\begin{abstract}
{We study the black body radiation in cavities of different geometry with different boundary conditions, and the formulas of energy spectral densities in films and rods are obtained, also the approximate energy spectral densities in cubic boxes are calculated. We find that the energy spectral densities deviate from Planck's formula greatly when the length(s) at some directions of the cavity can be compared to the typical wavelengths of black body radiation in infinite volume and the boundary conditions also affect the results. We obtain an asymptotic formula of energy spectral density for a closed cavity with Dirichlet boundary condition and find it still fails when the size of the cavity is too small.}
\bigskip


\end{abstract}

\maketitle


\maketitle

\section{Introduction}
\label{sec:Introduction}

In early days of quantum mechanics, black body radiation played an important role. 
In 1901, Planck first proposed
the concept of energy quantization when he studied black body radiation \cite{planck}, 
and this is just the origin of quantum mechanics.
It is well known that the Planck's formula (PF) for blackbody radiation states that the energy spectral density only depends on temperature, regardless of the shape and the size of the cavity. The derivation of this formula assumes that the size of the cavity 
is large enough and thus the density of state (the number of modes in a frequency interval) is independent of the geometry of it. Now it's natural to 
ask: Can PF correctly describe the blackbody radiation when the cavity size is no longer large enough?
To illustrate this clearly, let us consider the following case: at 300K, more than 99\% energy is carried by the modes whose wavelengths are larger than 3$\times 10^{-6}$ m according to Planck's formula. However, for example, in a cube whose length is shorter than 3$\times 10^{-6}$ m, the modes larger than 3$\times 10^{-6}$ m can't exist, so the spectral density in this cube will deviate greatly from the PF. Just as pointed out in Ref. \cite{rytov} that the PF is only applicable if $(\lambda/L)^3\ll\Delta\lambda/\lambda\ll 1$, where $L$ is the length of the cube and $\Delta\lambda$ is the wavelength interval. 
This clearly shows that when the size of the cavity is small enough, the PF is no longer suitable, so we need to develop new methods to study the blackbody radiation in small cavities.

After Planck's publication, the wave equation's eigenvalue distribution was investigated in Refs. \cite{weyl,asy,cou,asy1,asy2,asy3}, which is related to the correction of Planck's formula. By their method of asymptotic expansion, we can get some correction terms which relate to the geometry properties of the cavity as well as the boundary conditions. However, as indicated in Ref. \cite{pra}, we will show in this paper, this asymptotic expansion method has its inherent limitations, that is, the size of the cavity is limited small but still large enough compared to the wavelength of black body radiation. When the size of the cavity is  small enough, more exact but laborious method is needed.

In this work, we intend to study how geometry and boundary conditions affect blackbody radiation when the size of the cavity is small enough. This paper is organized as follows: In Sec II A, we study the energy spectral density in film and rod with periodic, antiperiodic and Dirichlet boundary conditions respectively. In Sec II B, we calculate the approximate energy spectral density in cubic box with periodic, antiperiodic and Dirichlet boundary conditions respectively. In addition, we compare the differences in energy spectral densities caused by geometry and boundary conditions. We also obtain an asymptotic formula of energy spectral density for a closed cavity with Dirichlet boundary condition and show in what range it is better than the PF and in what range it fails.

\begin{figure}                                                                                                                       
\begin{minipage}[t]{0.45\textwidth}                                                                           
\centering
\includegraphics[width=1\linewidth]{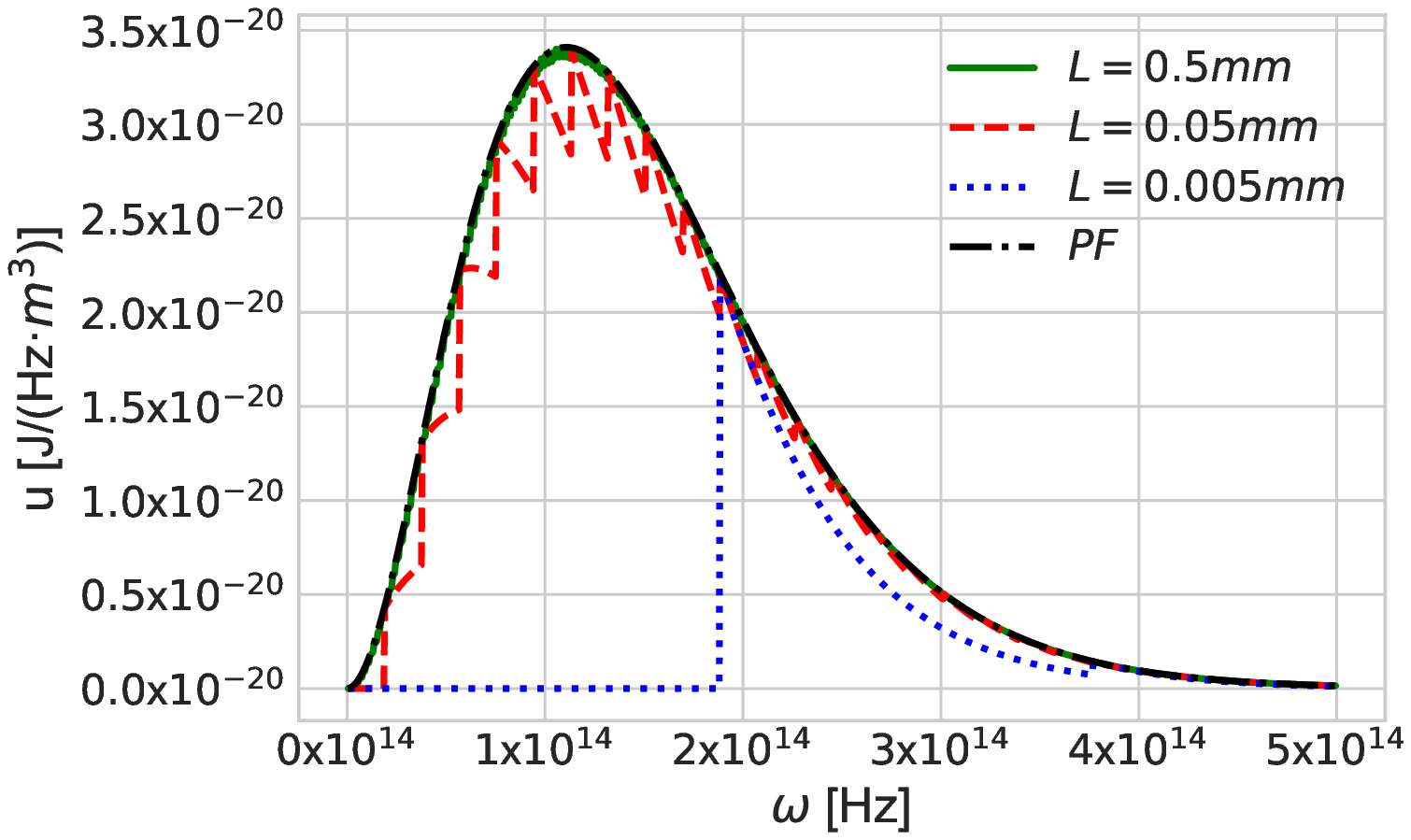} 	\\
\includegraphics[width=1\linewidth]{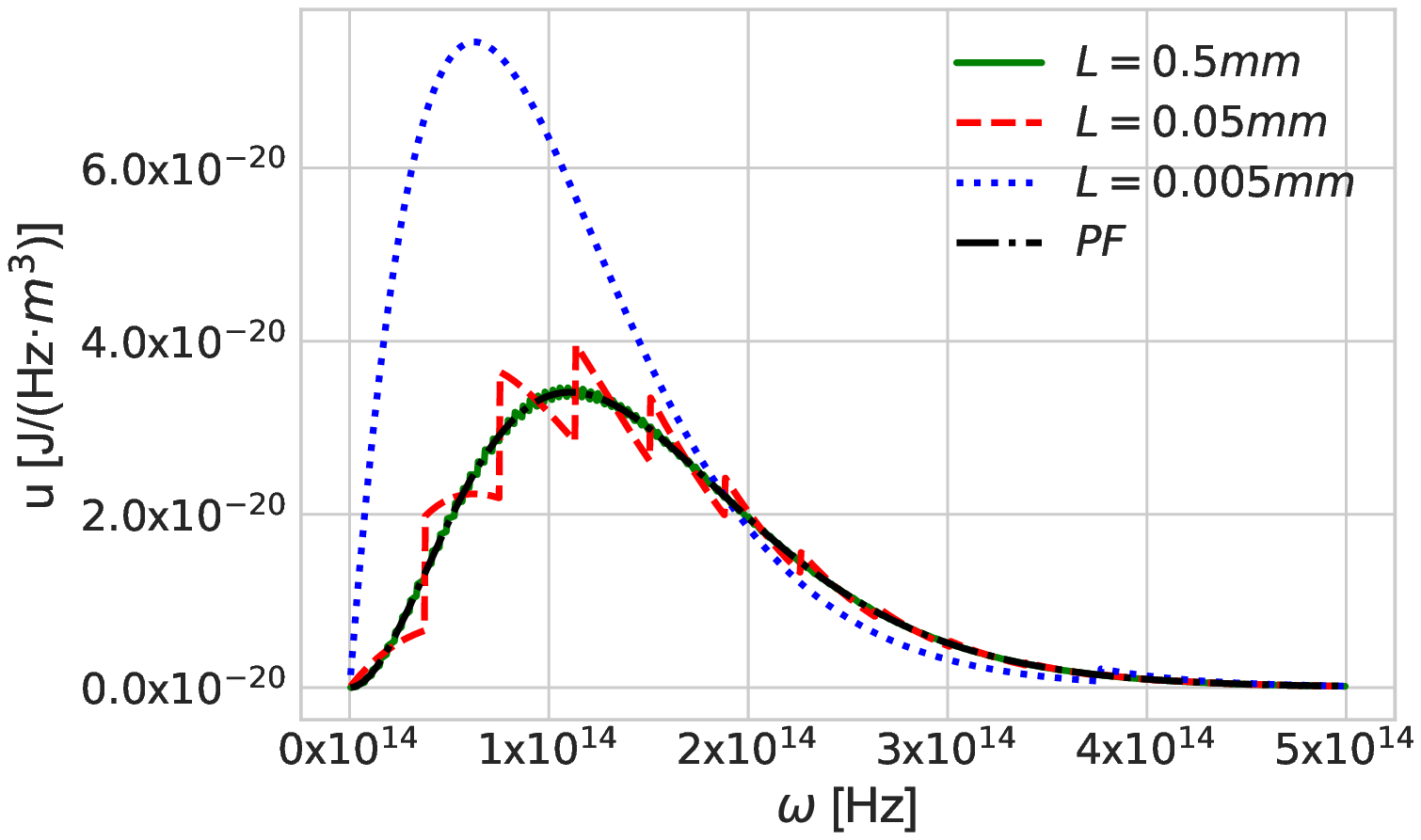} 	\\
\includegraphics[width=1\linewidth]{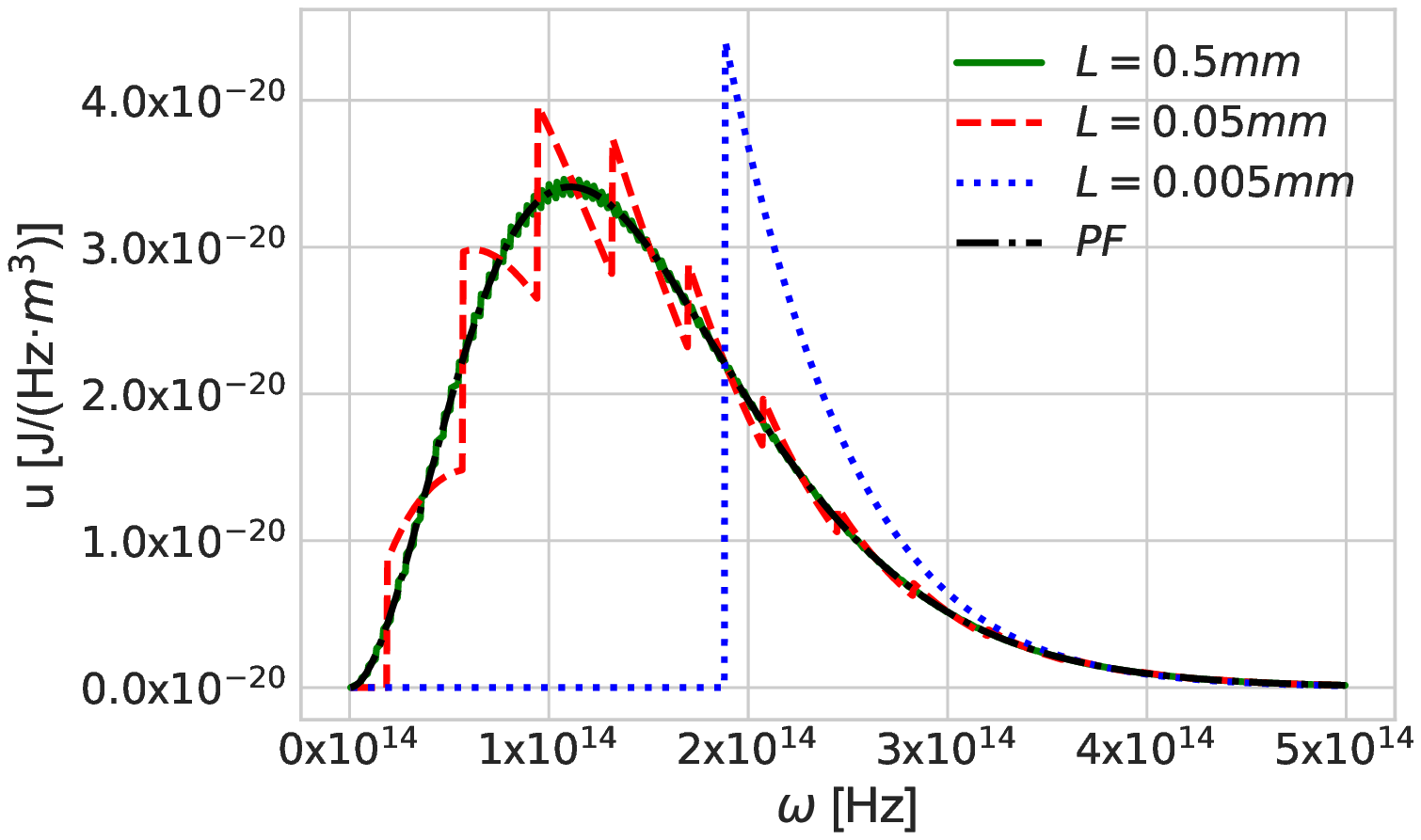} 
\caption{(Color online.) Energy spectral densities in a film with Dirichlet boundary condition (top panel), periodic boundary condition (central panel) and antiperiodic boundary condition (bottom panel) at 300 K. $L$ is the distance between the two surfaces of the film.}	
\end{minipage}
\end{figure}

\section{black body radiation in different cavities with different boundary conditions}
\label{sec:film}
\subsection{Film and Rod}
 First, let's review the derivation of PF. Consider an ensemble of oscillators in thermal equilibrium, by assuming the quantized values of an oscillator's energy, we obtain an oscillator with frequency $\omega$ has average energy $\overline{\epsilon}(\omega)=\hbar\omega[\mathrm{exp}(\hbar\omega/k_BT)-1]^{-1}$, where $k_B$ is the Boltzmann constant. Then we calculate the density of states (DOS) $g(\omega)\mathrm{d}\omega$ in the cavity, which means the number of modes in the interval $(\omega,\omega+\mathrm{d}\omega)$. If we assume that the size of the cavity is large enough, $\omega$ becomes continuous and we can get $g(\omega)={V}\omega^2/{\pi^2c^3}$, which is independent of the geometry of the cavity. Finally, the spatial energy spectral density (energy spectral density in unit volume, we will call it energy spectral density for short in this paper) can be written as 
\begin{equation}
u ( \omega , T ) \mathrm { d } \omega = \frac{\overline { \varepsilon }  g  ( \omega ) \mathrm { d } \omega}{V} = \frac { 1} {  \pi ^2 c ^ { 3 } } \frac { \hbar \omega ^ { 3 } \mathrm { d } \omega } { \mathrm { e } ^ { \hbar \omega / k T } - 1 },
\end{equation}	
which is the famous PF. It's notable here that Eq. (1) is independent of the geometry of the cavity, which is caused by                                                                                                                                                                                                                                                                                                                                                                                                                                                                                                                                                                                                                                                                                                                                                                                                                                                                                                                                                                                                                                                                           the geometric independence of $\overline{\epsilon}(\omega)$ and DOS. 
$\overline{\epsilon}(\omega)$ is derived from statistical considerations, and DOS is derived from geometry considerations, which is a good approximation only for a cavity with infinite volume. A proof for the validity of this approximation is given by Courant and Hilbert \cite{cou}.
  
Similar to Planck's derivation, in this work we assume that the small size of the cavity does not affect the Bose statistics of the photons, \emph{i.e}., we suppose the average energy formula $\overline{\epsilon}(\omega)=\hbar\omega[\mathrm{exp}(\hbar\omega/k_BT)-1]^{-1}$ still works.
This means that our correction of PF concentrates on the correction of DOS.
    
 Now, Let's derive the energy spectral density in a thin film, which can be regarded as two infinite plates at a distance. For the simplicity in later discussions, we first consider the eigenmodes of black body radiation in a box. $L_1, L_2$ and $L_3$ are the lengths of the box at $x, y$ and $z$ directions. If we take the periodic boundary conditions, the wave vector is given by
\begin{eqnarray}
k&=&\sqrt{k_x^2+k_y^2+k_z^2}\nonumber\\
&=& 2\pi \sqrt{\frac{n_1^2}{L_1^2}+\frac{n_2^2}{L_2^2}+\frac{n_3^2}{L_3^2}}, \nonumber \\
&& n_1,n_2,n_3=0,\pm1,\pm2,\dots ,
\end{eqnarray}
where $k_x, k_y$ and $k_z$ are the three components of the wave vector. In the case of film, we let $L_2$ and $L_3$ approach to infinite but $L_1$ be finite, thus $k_y$ and $k_z$ become continuous. It's easy to write the DOS 
\begin{equation}
g(k)\mathrm{d}k=\frac{L_2L_3}{\pi}\sum_{k_i}k\mathrm{d}k, \ \ \ |k_i|=\frac{2\pi |n_1|}{L_1}<k,\ n_1\in Z,
\end{equation}	
which can be written as 
\begin{equation}
g(k)\mathrm{d}k=\frac{L_2L_3}{\pi}(2[\frac{kL_1}{2\pi}]+1)k\mathrm{d}k,
\end{equation}	
where $[x]$ means integer-valued function. The energy spectral density is 
\begin{equation}
u(\omega,T)\mathrm{d}\omega=\frac{1}{\pi c^2L_1}\frac { \hbar \omega^2 } { \mathrm { e } ^ { \hbar\omega / k T } - 1 }(2[\frac{\omega L_1}{2c\pi}]+1)\mathrm{d}\omega.
\end{equation}
It depends on the distance between the two plates. When $L_1$ approaches to infinite, $2[{\omega L_1}/{2c\pi}]+1 \approx {\omega L_1}/{c\pi}$, Eq. (5) reduces to the PF.

If the antiperiodic boundary condition is taken, we can get 
\begin{equation}
u(\omega,T)\mathrm{d}\omega=\frac{2}{\pi c^2L_1}\frac { \hbar \omega^2 } { \mathrm { e } ^ { \hbar\omega / k T } - 1 }[\frac{\omega L_1}{2c\pi}+\frac{1}{2}]\mathrm{d}\omega.
\end{equation}
Similarly, in Dirichlet boundary condition, we have
\begin{equation}
u(\omega,T)\mathrm{d}\omega= \frac{1}{\pi c^2L_1}\frac { \hbar \omega^2 } { \mathrm { e } ^ { \hbar\omega / k T } - 1 }[\frac{\omega L_1}{c\pi}]\mathrm{d}\omega.
\end{equation}
It's easy to find the results with different boundary conditions are different, especially when $L_1$ is small compared to the typical wavelengths. Fig. 1 shows the energy spectral densities of films at different $L_1$ with different boundary conditions. For antiperiodic and Dirichlet boundary conditions, there's a cut-off frequency. But for periodic boundary condition, there's not such a cut-off frequency and thus the energy density tends to infinite as $L_1$ tends to 0. For a slim rod with periodic boundary condition, there is also such "infrared divergence". The authors of Ref. \cite{pra} dealt with the problem by claiming that every experimental apparatus has a resolution $\Delta E$ and we need not to count the modes whose $\omega<\Delta E/\hbar $. For Dirichlet boundary condition, energy densities at every frequency are not higher than that of the PF's results, which can be easily seen from Eq. (7), because $[x]\leq x$. But for periodic and antiperiodic boundary conditions, energy densities at some frequencies are higher than PF's results. When $L_1$ is large enough compared to the typical wavelengths, the energy spectral densities with different boundary conditions all coincide with the PF.

Similar to the derivation above, we derive the energy spectral density of a slim rod. In Eq. (2), we let $L_3$ approach to infinite but $L_1, L_2$ be finite, it's also easy to write the DOS
\begin{equation}
\begin{split}
 g(k)\mathrm{d}k=\frac{2L_3}{\pi}\sum_{k_1^2+k_2^2<k^2}\frac{k\mathrm{d}k}{\sqrt{k^2-(k_1^2+k_2^2)}},\\
(k_1,k_2)=(\frac{2n_1\pi}{L_1},\frac{2n_2\pi}{L_2}),\ n_1,n_2 \in Z.
\end{split}
\end{equation}
 The energy spectral density is 
\begin{eqnarray}
	u(\omega,T)\mathrm{d}\omega &=& \frac{2}{\pi c^2L_1L_2}\frac { \hbar \omega^2 } { \mathrm { e } ^ { \hbar\omega / k T } - 1 }\times \nonumber\\
	&&\sum_{k_1^2+k_2^2<\frac{\omega^2}{c^2}}\frac{\mathrm{d}\omega}{\sqrt{\frac{\omega^2}{c^2}-(k_1^2+k_2^2)}},\nonumber\\
	 (k_1,k_2)&=&(\frac{2n_1\pi}{L_1},\frac{2n_2\pi}{L_2}),\ n_1,n_2 \in Z.
\end{eqnarray}
In antiperiodic boundary condition case, we just need to replace $(k_1, k_2)$ by
\begin{eqnarray}
(k_1,k_2)&=&(\frac{2(n_1+1/2)\pi}{L_1},\frac{2(n_2+1/2)\pi}{L_2}),\ n_1,n_2 \in Z.
\nonumber\\
\end{eqnarray}
in Eq. (9).
Also, in Dirichlet boundary condition case, we replace $(k_1, k_2)$ by
\begin{eqnarray}
(k_1,k_2)&=&(\frac{n_1\pi}{L_1},\frac{n_2\pi}{L_2}),\ n_1,n_2 \in N^*.
\end{eqnarray}
in Eq. (9). Fig. 2 shows the energy spectral densities in rods of different $L_1$ ($L_2=L_1$) with different boundary conditions. It is found that the energy spectral densities reduce to PF when $L_1$ is large enough. Similar to the case of film, when $L_1$ is small compared to the typical wavelengths, the energy spectral densities deviate from PF greatly. The boundary conditions also influence the results. It is noticed for all these three boundary conditions, the energy densities at some frequencies are higher than PF's results. We also notice for a rod, the curves of spectrums oscillate more violent than that of a film.

An important thing we note is that so long as the length(s) in the constrained direction(s) is small enough, the spectrum will deviate from the PF greatly even at high frequencies where the PF is supposed to be exact. 

Just as shown above we discussed the effect of small size on energy spectral density in the case of film and rod, in the next section, we will study the effect of small size on the blackbody radiation in a cubic box.
\begin{figure}
\centering
\begin{minipage}[b]{0.45\textwidth}
\includegraphics[width=1\linewidth]{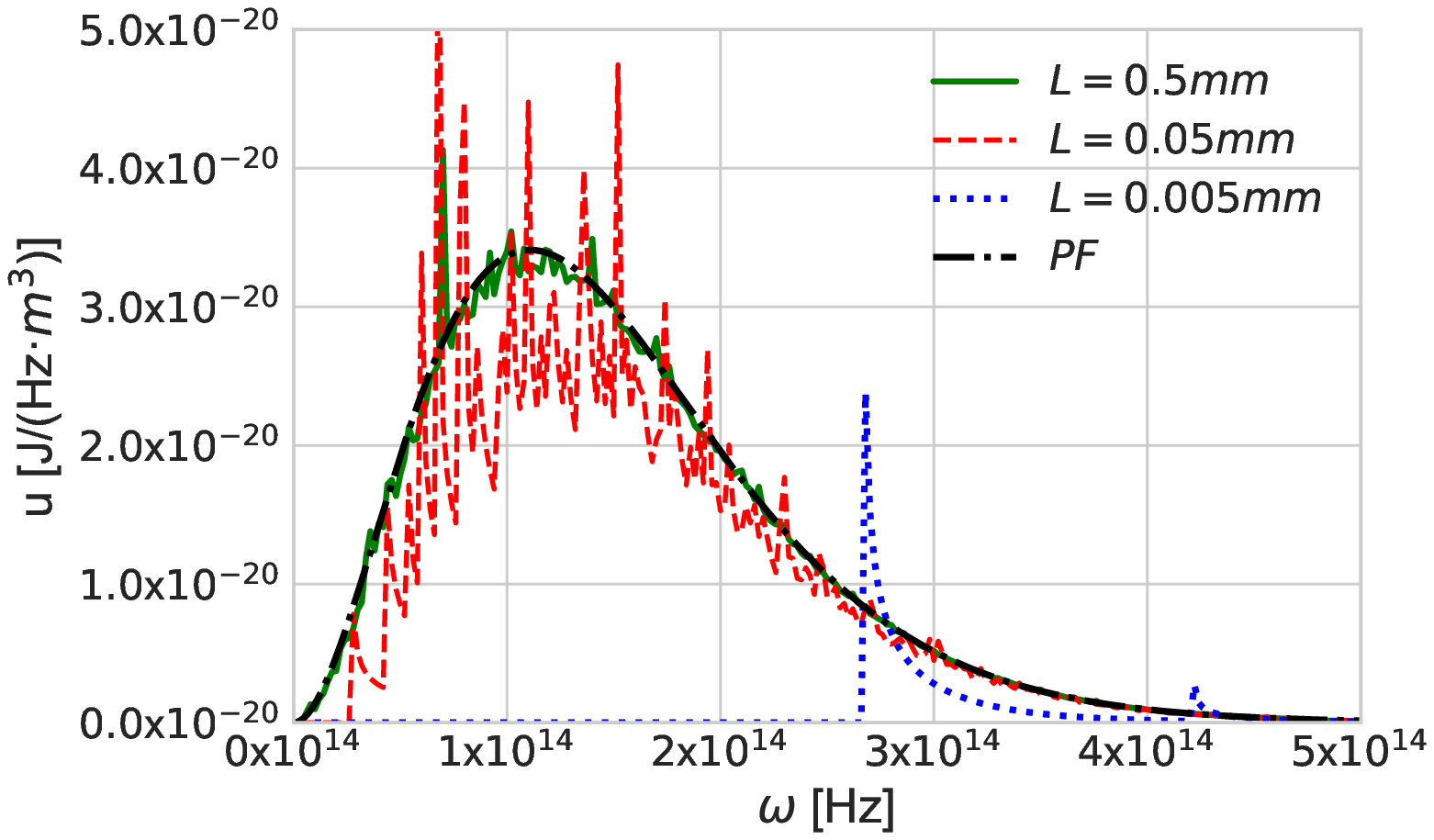} \\
\includegraphics[width=1\linewidth]{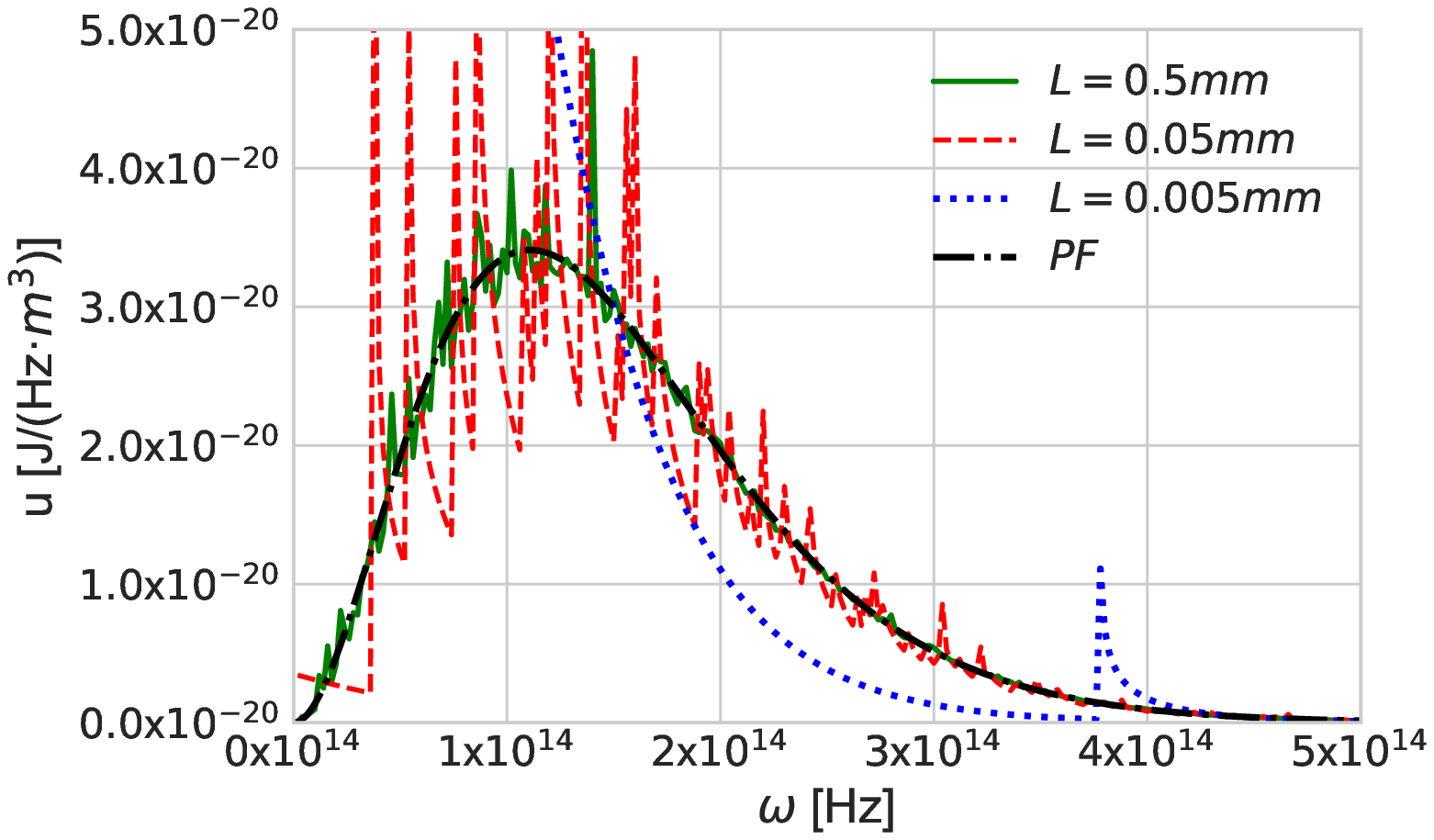} \\
\includegraphics[width=1\linewidth]{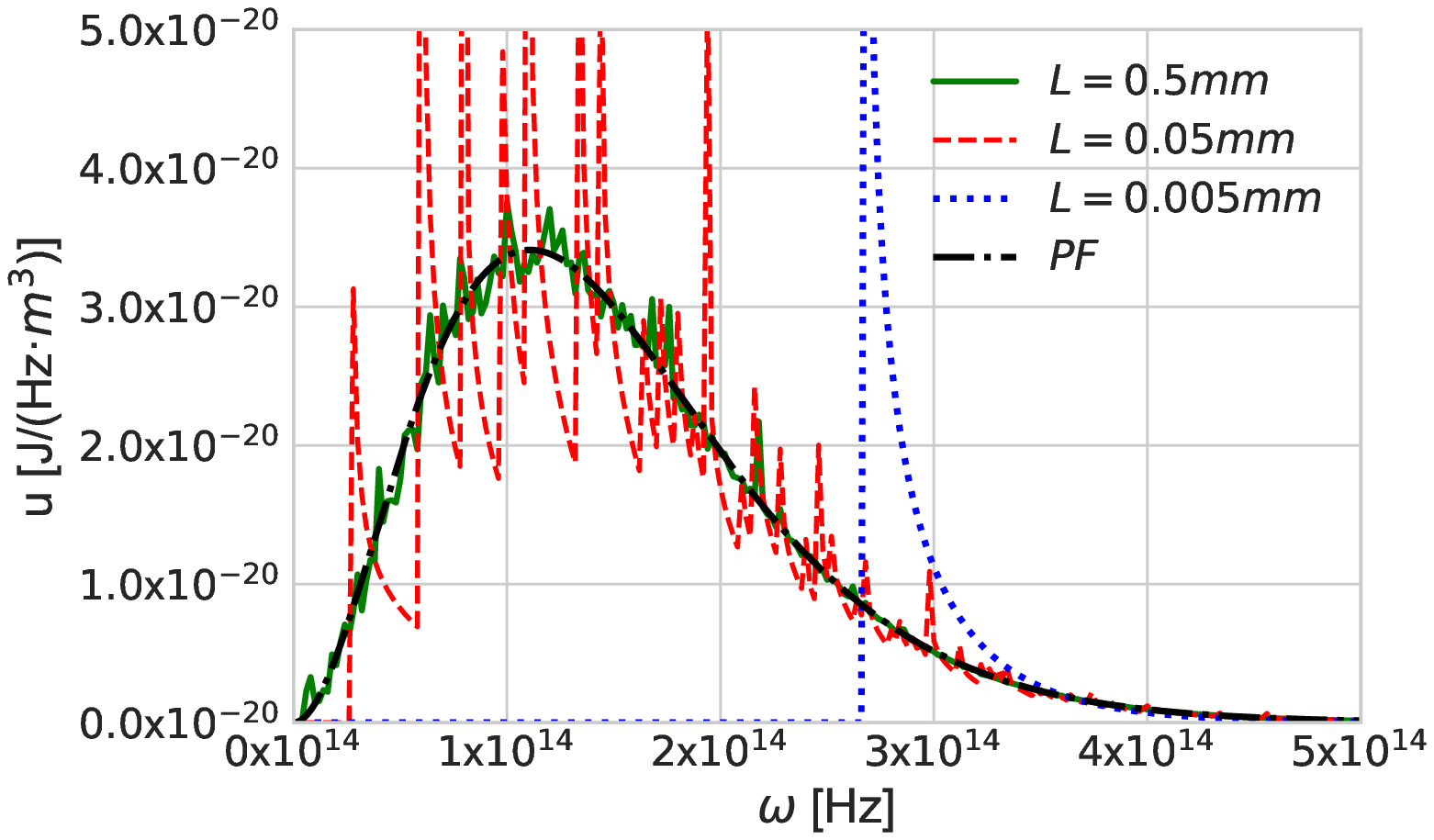} \\
\end{minipage}
\caption{(Color online.) Energy spectral densities in a rod with Dirichlet boundary condition (top panel), periodic boundary condition (central panel) and antiperiodic boundary condition (bottom panel) at 300 K. Here $L_1=L_2=L$ is taken.}
\end{figure}

\subsection{Box}
\label{sec:Box}
Different from the cases of film and rod, we can not define DOS for a cavity with finite volume because of the discrete values of frequencies, so we also can't define the energy spectral density for such systems. But we can define an approximate energy spectral density. If we
 divide the frequency into many small intervals with the length of $\Delta\omega$, the energy spectral density on $(\omega,\omega+\Delta\omega)$ can be defined as
\begin{equation}
u(\omega,T)=\sum_{\omega<\omega_i<\omega+\Delta\omega}\frac{N(\omega_i)}{V\Delta\omega}\frac { \hbar \omega_i } { \mathrm { e } ^ { \hbar\omega_i / k T } - 1 },
\end{equation}
where $\omega_i$ are the allowed discrete values of $\omega$ and $N(\omega)$ is the number of modes of frequency $\omega$.  When the number of modes is great enough in each interval, the approximate energy spectral density appears to be continuous and inert to the change of $\Delta\omega$, as long as $\Delta\omega$ is still small enough. This definition is similar to the definition given by 
Ref. \cite{pra}, but our $\Delta \omega$ is a constant once specified.

\begin{figure}
\begin{minipage}[b]{0.45\textwidth}
\includegraphics[width=1\linewidth]{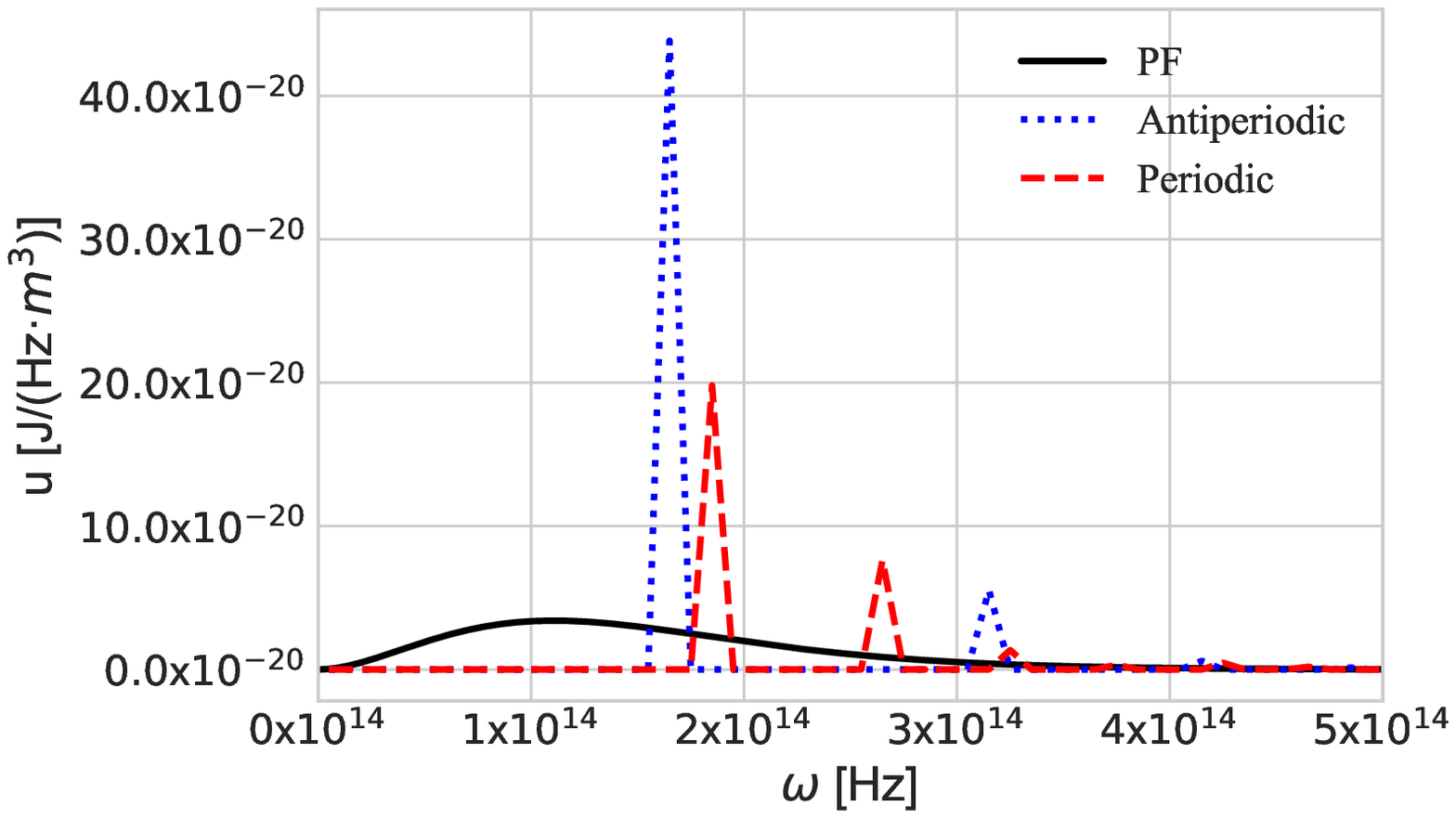} \\
\includegraphics[width=1\linewidth]{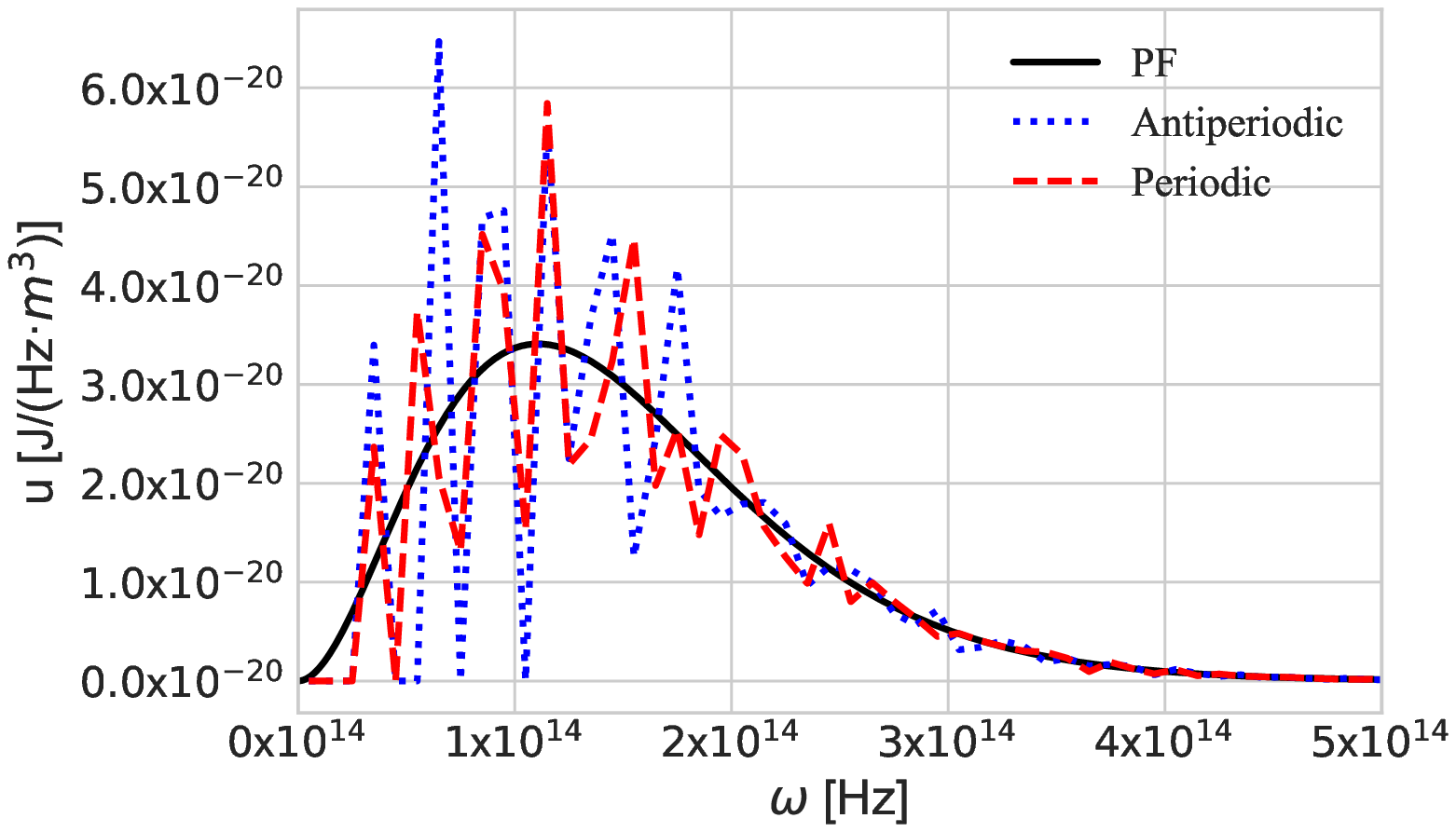} \\
\includegraphics[width=1\linewidth]{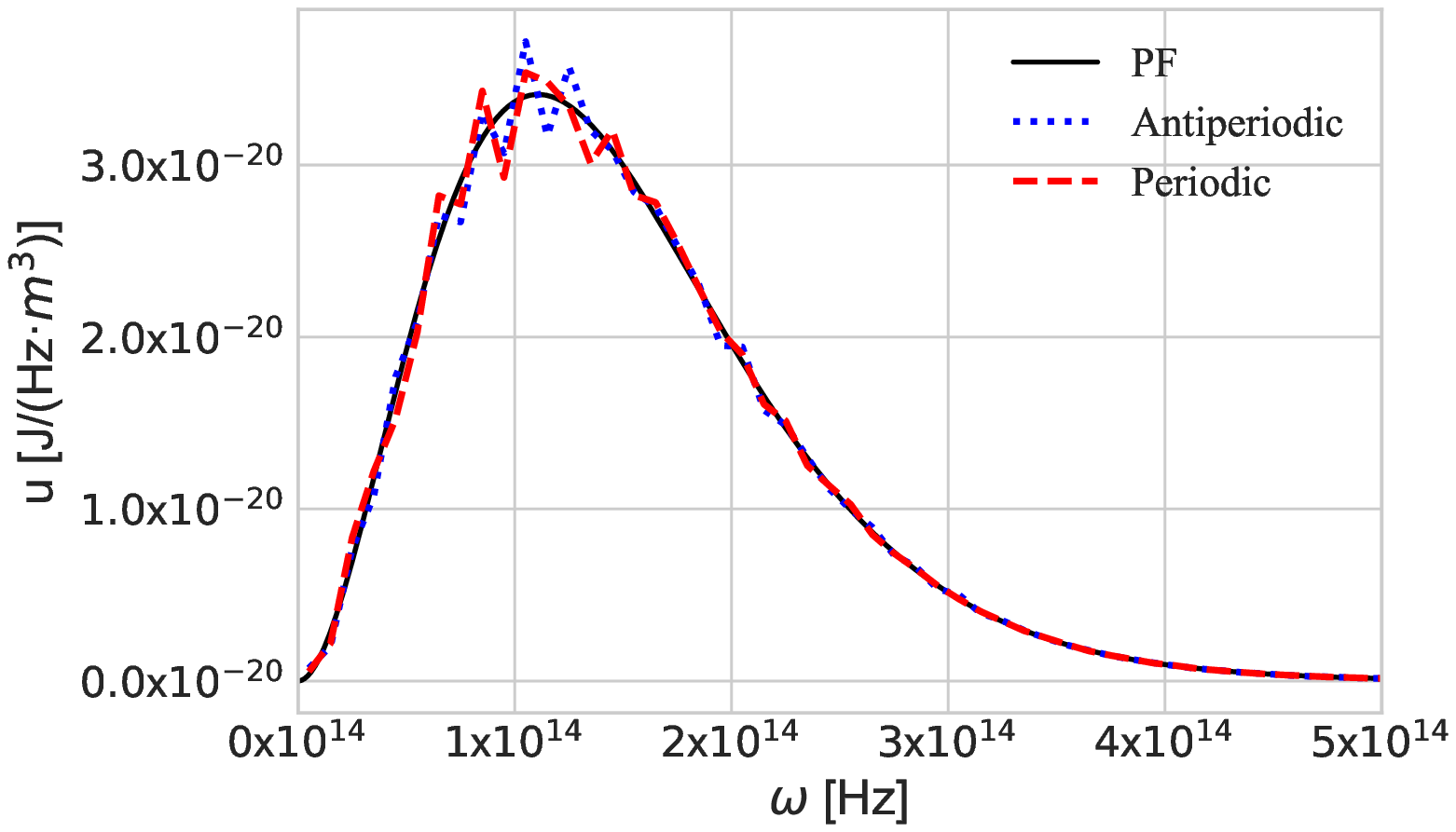} \\
\end{minipage}
\caption{(Color online.) Energy spectral densities in a cubic box with periodic boundary condition (red) and antiperiodic boundary condition (blue) at 300 K, $\Delta\omega=10^{13}$ Hz. The cavity sizes are with different values at  $L_1$=0.01 mm  in top panel, $L_1$=0.05 mm in central panel and  $L_1$=0.2 mm in bottom panel respectively.}
\end{figure}

For a cubic box with the periodic boundary condition, the wave vector takes the values given in Eq. (2). We run over all integers to get $N(\omega)$ and then calculate the approximate energy spectral density. Antiperiodic boundary condition and Dirichlet boundary condition cases are obtained in the same way. Fig. 3 shows the approximate energy spectral densities of a cubic box at different length $L$ with periodic and antiperiodic boundary conditions. We can see that when $L$ is large enough, the approximate energy spectral densities coincide with PF but when $L$ is small enough, they deviate from the PF greatly. When $L$ is small, the oscillation of the curves indicates that there are only a few modes in each interval $\Delta\omega$, which means that the spectral density cannot be considered as continuous anymore, whereby the concept of energy spectral density is not suitable. It is amazing to find that if the length of the cubic box is small enough, the energy density in the box will become very small compared with that in infinite volume.

\begin{figure*}
\centering
\subfigure[ box, $\Delta\omega=10^{13}$ Hz]{
\begin{minipage}[b]{0.45\textwidth}
\includegraphics[width=1\linewidth]{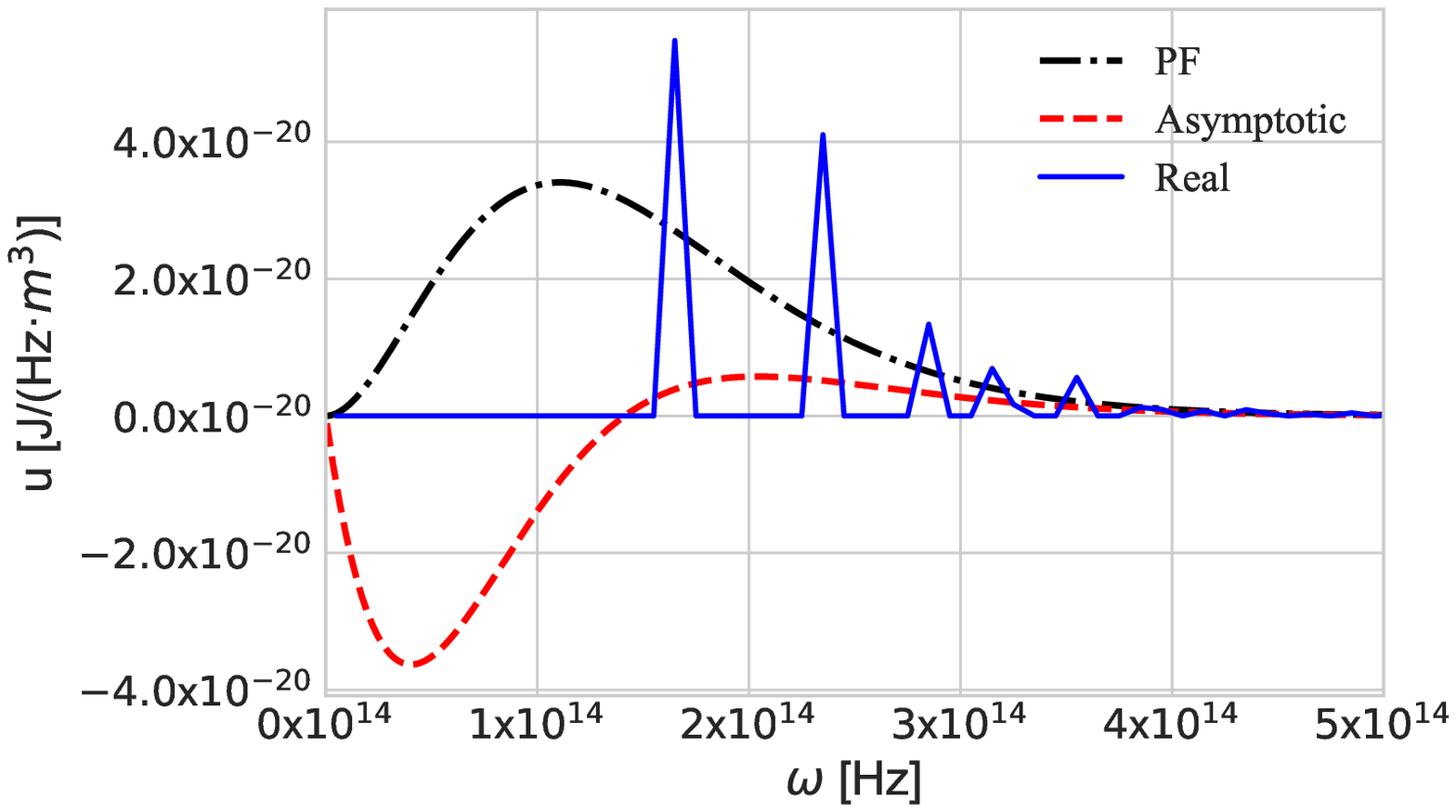} \\
\includegraphics[width=1\linewidth]{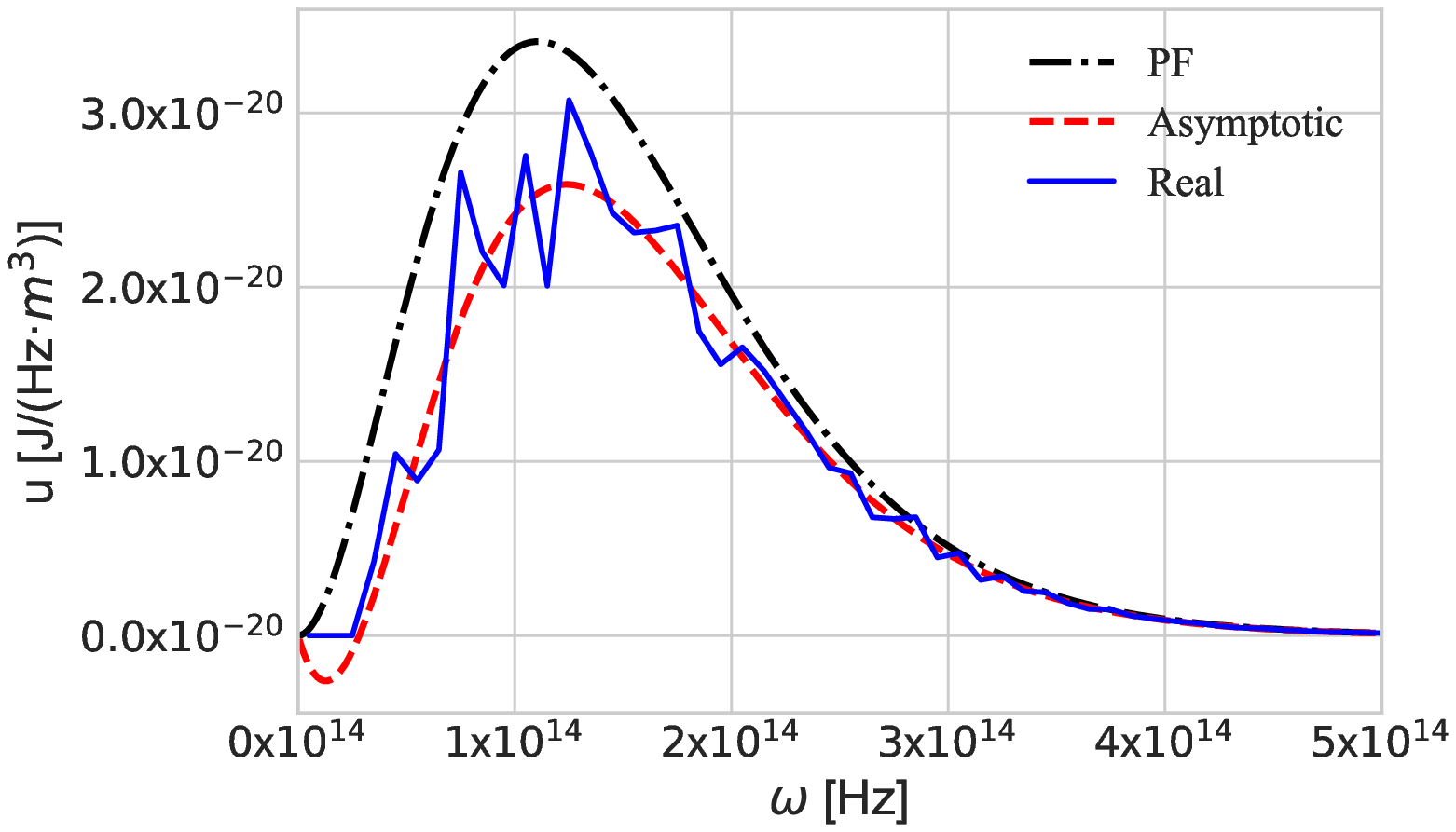} \\
\includegraphics[width=1\linewidth]{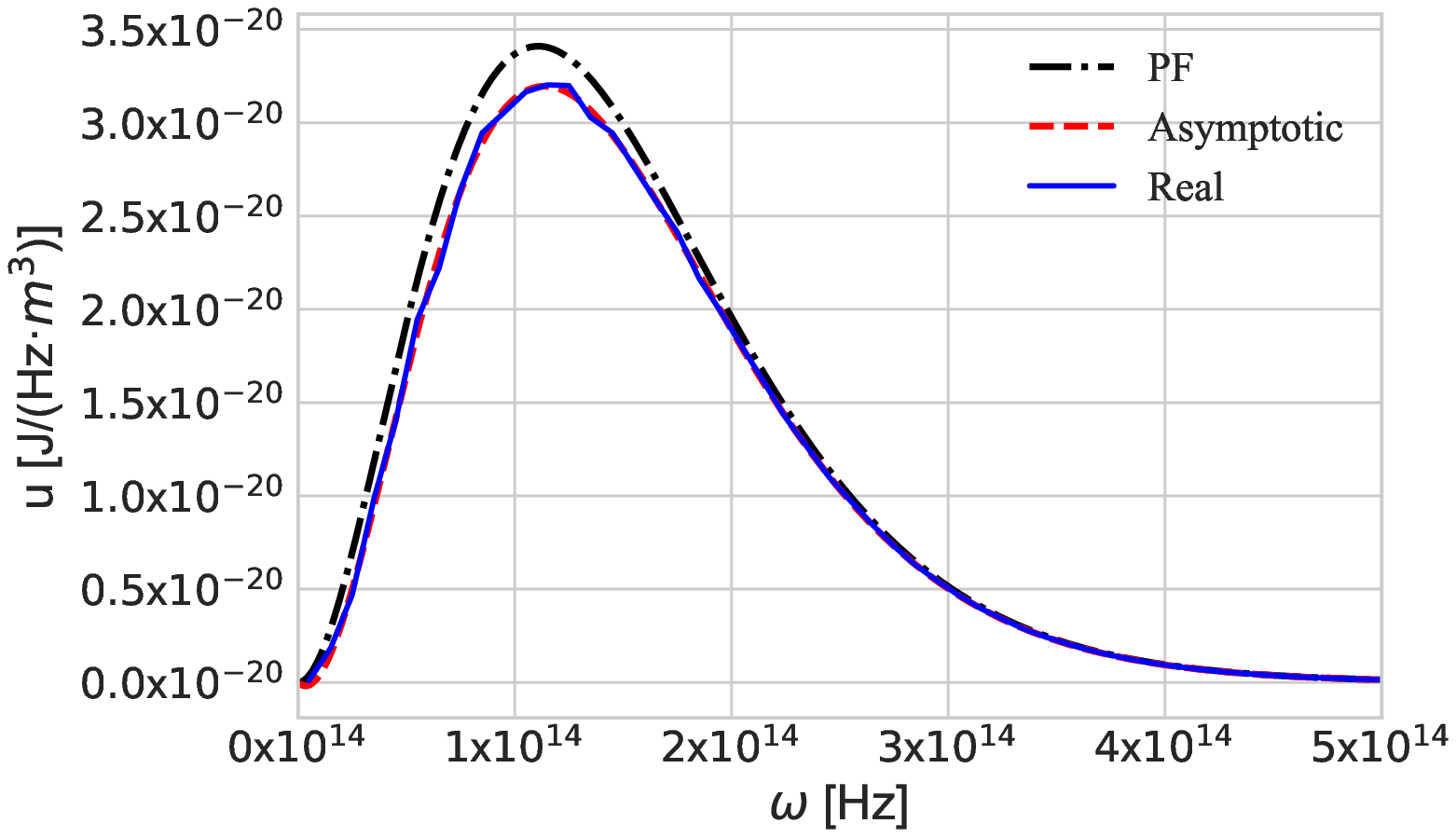} \\
\end{minipage}}
\subfigure[ sphere, $\Delta\omega=10^{13}$ Hz]{
\begin{minipage}[b]{0.45\textwidth}
\includegraphics[width=1\linewidth]{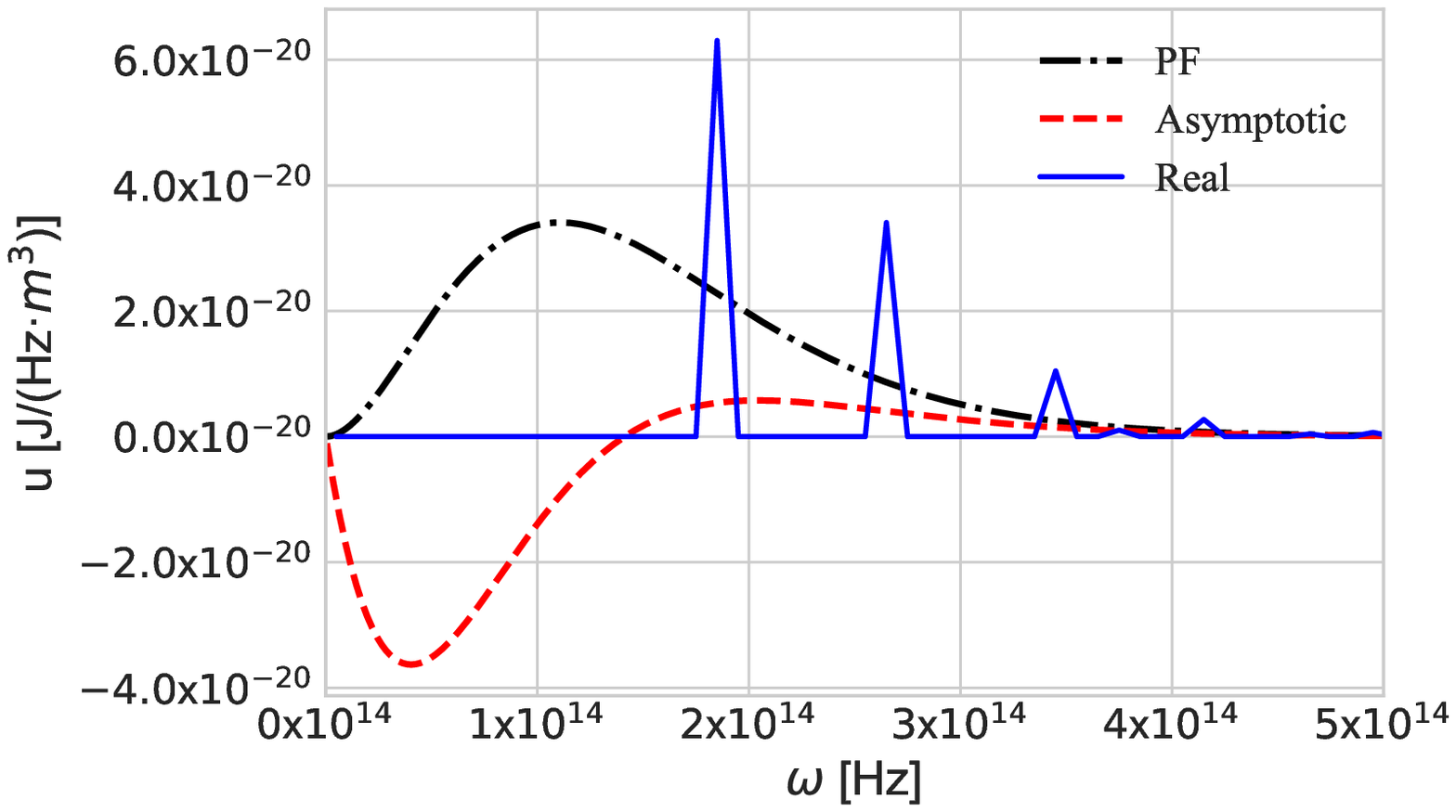} \\
\includegraphics[width=1\linewidth]{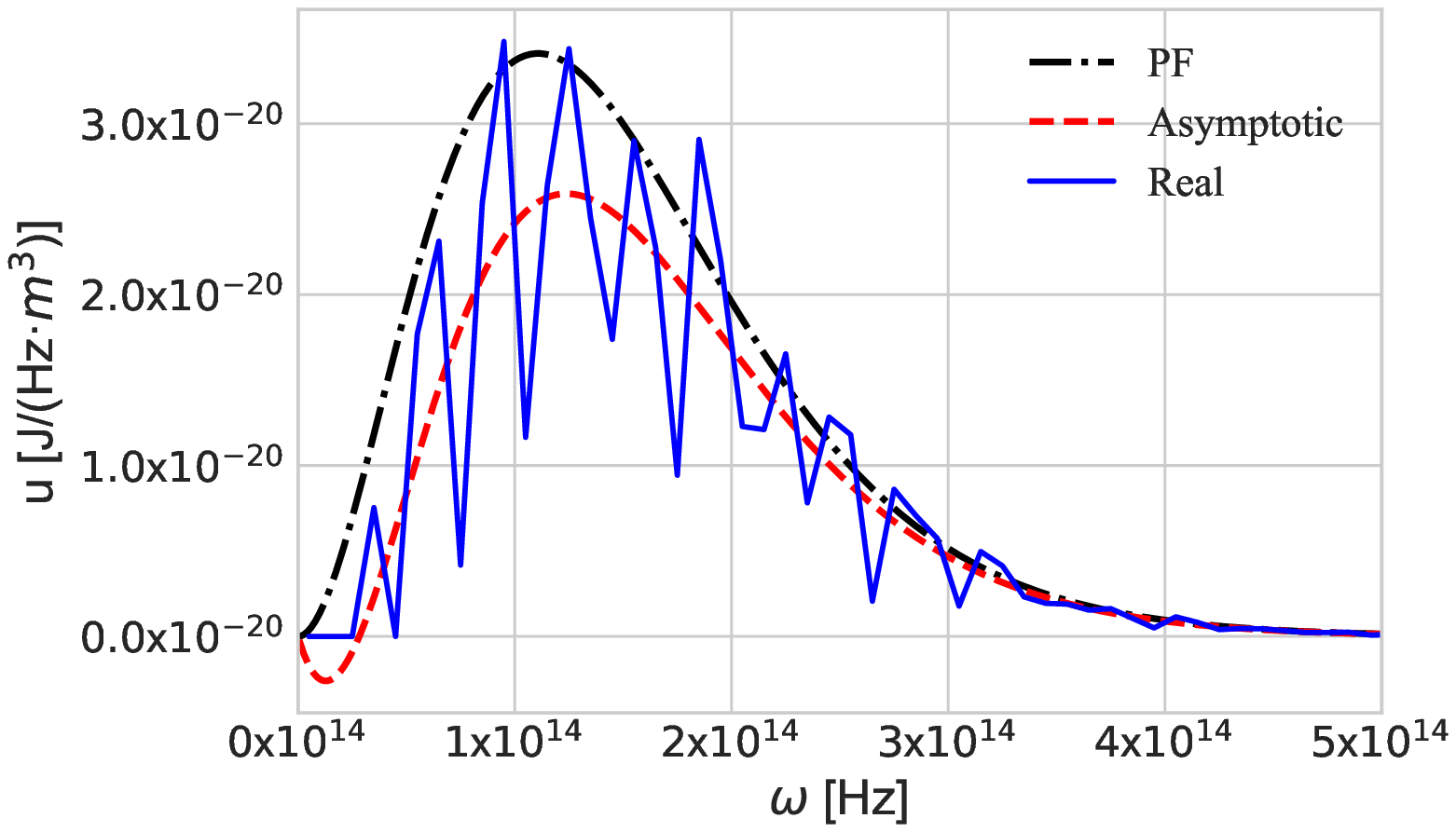} \\
\includegraphics[width=1\linewidth]{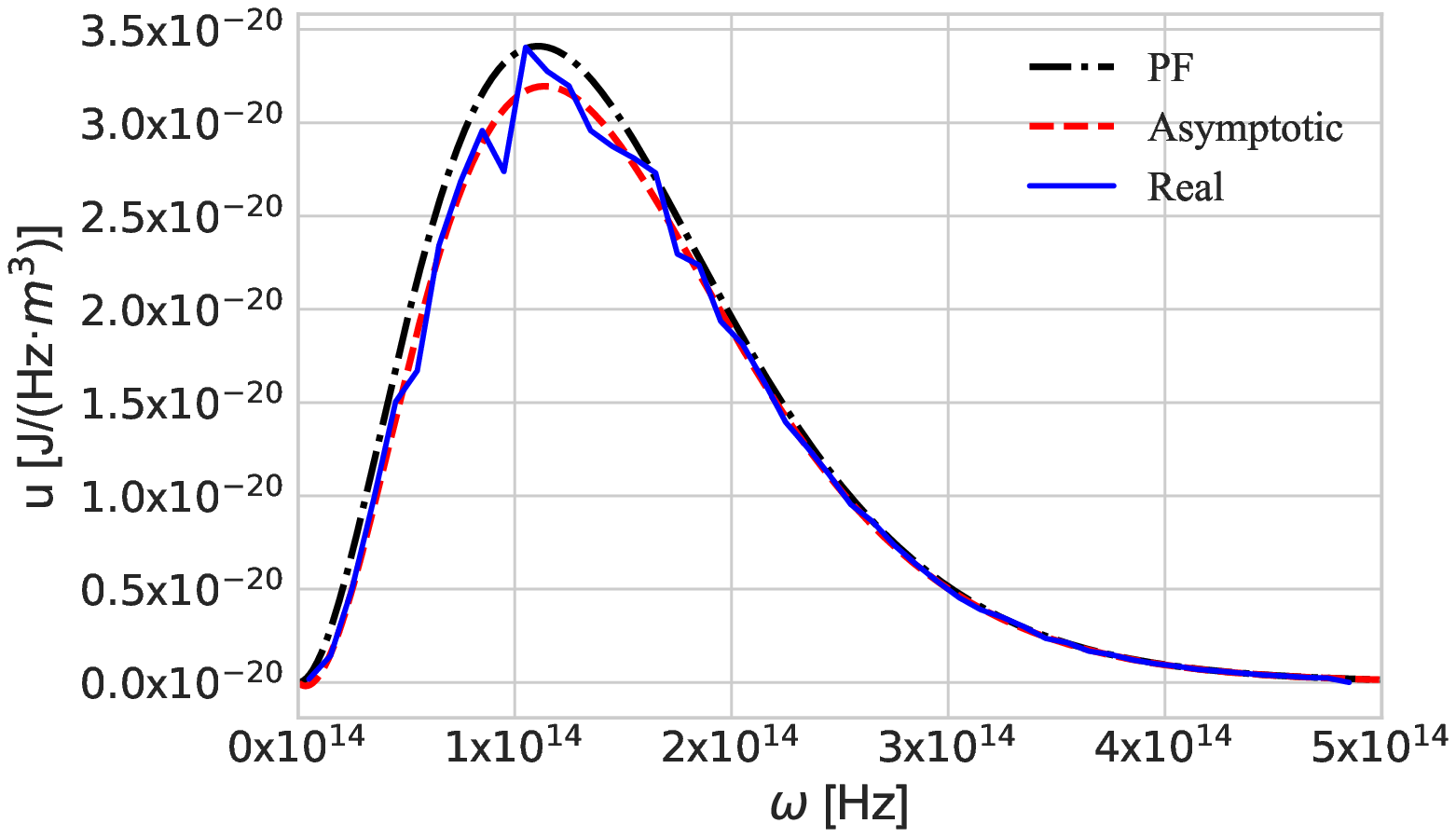} \\
\end{minipage}
}
\caption{(Color online.) Energy spectral densities  in different geometric cavities with Dirichlet boundary condition at 300 K. The left       panel is in a cubic box with  $L_1$=0.01 mm (top panel), $L_1$=0.05 mm (central panel), and  $L_1$=0.2 mm (bottom panel) respectively. The right panel is in a sphere with d=0.01 mm (top panel), d=0.05 mm (central panel),  and d=0.2 mm (bottom panel) respectively.}
\end{figure*}
Now we want to know when the asymptotic formula (correction formula by the method of asymptotic expansion) will work well and fail. From the result given in Ref. \cite{math}, we obtain an asymptotic formula with Dirichlet boundary condition for a closed cavity with an arbitrary shape,
\begin{equation}
u_{asy}(\omega,T)\mathrm{d}\omega=(\frac{\hbar\omega^3}{\pi^2c^3}-\frac{A}{V}\frac{\hbar\omega^2}{4\pi c^2}+\frac{M}{V}\frac{\hbar\omega}{3\pi^2c})\frac{1}{{ \mathrm { e } ^ { \hbar\omega / k T } - 1 }}\mathrm{d}\omega,
\end{equation}
where $A$ is the surface area of the cavity, $M$ is the surface integral of mean curvature. $M=\int_S{\frac{1}{2}(\kappa_1+\kappa_2)\mathrm{d}S}$ with $\kappa_1$ and $\kappa_2$ the principle curvatures at the surface. Fig. 4 shows the comparison between the PF, Eq. (13) and the approximate energy spectral density by Eq. (12) for a cubic box and a sphere with Dirichlet boundary condition. We can see when the length of the cavity is so small that the spectral density becomes discrete, both PF and (13) fail and (13) even gives meaningless minus results. When the cavity is large enough, the three coincide each other. When $L$ is neither extremely small nor large, Eq. (13) is more close to the real energy spectral density than PF. 

As shown in Figs. (2-4), when the cavity size is large enough, the blackbody radiation spectrums obtained by three different boundary conditions tend to the standard PF, which we can expect in advance.
However, when the cavity size is small enough, the blackbody radiation spectrums obtained with different boundary conditions are completely different. Then, a natural problem arises, that is, which boundary condition is true? Regarding this issue, our views are as follows: When we consider the blackbody radiation in a small cavity , since the photons cannot escape  from the cavity, it is theoretically necessary to use Dirichlet boundary condition. Of course, this requires further testing by the experiment.

\section{Conclusion}
\label{sec:Conclusion}
In this paper, we investigated the geometry effect on blackbody radiation with different boundary conditions. Here we summarize the main conclusions of this study (include some common knowledge):

	(1) We derived the formula of blackbody radiation for a film and a rod with different boundary conditions.

	(2) The geometry effect of black body radiation is great when the length(s) at constrained direction(s) of the cavity is not large compared to the typical wavelengths of black body radiation. 

	(3) Planck's formula can not describe black body radiation properly in a very small cavity, neither does asymptotic formula, which is better than PF when the volume is finite and not too small.

	(4) The boundary conditions also influence blackbody radiation, especially when in small cavities.

	(5) The concept of energy spectral density is not suitable for a closed cavity, especially when the size of the cavity is close to the wavelengths of blackbody radiation.

Finally, we stress that in this paper we still assume that the small size does not affect the Bose statistics, which obviously needs to be tested experimentally, and deserves further research. 
In fact, some experiments have recently suggested that small volume effect may affect quantum statistical distribution. For example, as shown in Ref. \cite{BEC}, no more than ten photons show Bose-Einstein condensation in a small cavity.  This indicates that new physics may exist in the extremely small volume.
\section*{Acknowledgments}

This work is supported in part by the National Natural Science Foundation of China (under Grants No. 11475085, No. 11535005, and No. 11690030) and by Nation Major State Basic Research and Development of China (2016YFE0129300).

\bibliography{reference}

\begin{thebibliography}{4}

\bibitem{planck}  M.Plank, Ueber das Gesetz der Energieverteilung im Normalspectrum, Annalen der Physik. 4th series (in German). 4 (3), 553-563 (1901).

\bibitem{pra} A. Reiser and L. Sch\" achter, Phys. Rev. A 87, 033801 (2013).
\bibitem{rytov} S. M. Rytov, Y. A. Kravtsov, and V. I. Tatarskii, Principles of Statistical Radiophysics, Vol 3: Elements of Random Fields
(Springer-Verlag, Berlin, 1989). 
\bibitem{weyl} H. Weyl, Math. Ann. 71, 441 (1912). 

\bibitem{math} R. T. Waechter, Proc. Cambridge Philos. Soc. 72, 439-447 (1972).

\bibitem{cou} R. Courant and D. Hilbert, Methods of Mathematical Physics (Wiley, New York, 1989), Chap. VI, Sec. 4.

\bibitem{asy1} T. Carleman, in Proceedings of the Eighth Scandinavian Mathematics Congress, Stockholm (Ohlsson, Lund, 1935), p.34.
\bibitem{asy2} \AA . Pleijel, Commun. Pure Appl. Math. 3, 1 (1950).
\bibitem{asy3} F. H. Brownell, Pacific J. Math. 5, 483 (1955).
\bibitem{asy} H. P. Baltes and E. R. Hilf, Spectra of Finite Systems (Bibliographisches Institut, Mannheim, 1976).
\bibitem{BEC} B. T. Walker, L. C. Flatten, H. J. Hesten, F. Mintert, D. Hunger, Aur\'{e}lien A. P. Trichet, J. M. Smith, and R. A. Nyman, Nat. Phys. 14, 1173 (2018).
\end{thebibliography}
\end{document}